\documentstyle[multicol,prl,aps,epsf,epsfig]{revtex} 

\begin{document}

\title{Spin dependent observable effect for free particles using the arrival time distribution}

\author{ Md. Manirul Ali\footnote{mani@bose.res.in}$^1$,
A. S. Majumdar\footnote{archan@bose.res.in}$^1$, 
Dipankar Home$^2$, and~Shyamal Sengupta$^3$}

\address{$^1$S. N. Bose National Centre for Basic Sciences,
Block JD,
Sector III, Salt Lake, Calcutta 700098, India}
 
\address{$^2$Department of Physics, Bose Institute, Calcutta
700009, India}

\address{$^3$Jadavpur University, Calcutta 700032, India}
 
\maketitle
 
\begin{abstract}

 The mean arrival time of free particles is computed using the quantum
probability current. This is uniquely determined in the non-relativistic
limit of Dirac equation, although the Schroedinger probability
current has an inherent non-uniqueness. Since the Dirac probability
current involves a spin-dependent term, an arrival time distribution
based on the probability
current shows an observable spin-dependent effect, even for free
particles. This
arises essentially from relativistic quantum dynamics, but persists
even in the non-relativistic regime.

PACS number(s): 03.65.Bz
 
\end{abstract}
 
\begin{multicols}{2}

\section{Introduction}

The treatment of time in quantum mechanics is a much debated 
question. A testimony to this is the proliferation of recent papers
\cite{muga} on the problems of tuneling time, decay time, dwell time and
the arrival time. In this paper we are specifically concerned with the
issue of arrival time\cite{muga2}.

In classical mechanics, a particle follows a definite trajectory;
hence the time at which a particle reaches a given location is a well
defined concept. On the other hand, in standard quantum mechanics,
the meaning of arrival time is rather problematic. Indeed, there exists
an extensive literature on the treatment of arrival time distribution
in quantum mechanics\cite{leavens}. A straightforward procedure would be
to try to construct a self-adjoint operator for the arrival time in quantum
mechanics which is conjugate to the Hamiltonian, but then it is seen that 
the operator does not have a basis of orthogonal eigenstates\cite{rovelli}.

Using the Born interpretation, $|\psi({\bf x},t_1)|^2$,~$|\psi({\bf x},t_2)|^2$... give the 
position probability distributions at different instants $t_1$, $t_2$.... Now, the question posed
is that if we fix the positions at ${\bf x}$=${\bf X}_1$,~${\bf X}_2$...,
can the functions $|\psi({\bf X}_1,t)|^2$,~$|\psi({\bf X}_2,t)|^2$...give the time probability
distributions at different positions ${\bf X}_1$,~${\bf X}_2$...?
Note that if at any instant $t=t_i$, $\int^{+\infty}_{-\infty}|\psi({\bf x},t=t_i)|^2 d^3x=1$,
the probability of finding the particle anywhere at that instant is unity. But if we fix the 
position at, say, ${\bf x}={\bf X}_1$ and t is varied, the value of the integral 
$\int^{\infty}_{0}|\psi({\bf x}={\bf X_1},t)|^2$dt $\ne$ 1. 
In this case what may be pictured is that at a given point, say, ${\bf X}_1$
the relevant probability changes with time and
this change of probability is governed by the following continuity equation which
suggests a ``flow of probability''

\begin{equation}
\frac{\partial}{\partial t}|\psi({\bf x},t)|^2 + {\bf \nabla}.{\bf J}({\bf x},t)=0
\end{equation}
where ${\bf J}({\bf x},t)$=$\frac{i\hbar}{2m}(\psi {\bf \nabla} \psi^{\ast}-\psi^{\ast}{\bf \nabla} \psi)$ is the probability current density.

Different approaches for analysing the problem of arrival time distribution
have been suggested using the path integrals and positive-operator-valued
measures \cite{others}. Delgado and Muga\cite{delgado1} proposed an interesting approach 
by constructing a self-adjoint operator having dimensions of time which is
relevant to the arrival time distribution, but then its conjugate Hamiltonian
has an unbounded spectrum. The implications of this approach have been studied
in detail by Delgado\cite{delgado2}.  

In this paper we take recourse to the definition of arrival time distribution
in terms of the quantum probability current density 
${\bf J}({\bf x}={\bf X},t)$.
Interpreting the equation of continuity in terms of the 
physical probability flow, the Born interpretation for the modulus of the
wave function and its time derivative seems to imply that
the mean arrival time of the particles reaching a detector located at ${\bf X}$
is given by

\begin{eqnarray}
\bar {\bf \tau}=\frac{\int^{\infty}_{0}|{\bf J}({\bf x}={\bf X}, t)| t  dt}
{\int^{\infty}_{0}|{\bf J}({\bf x}={\bf X}, t)| dt}
\end{eqnarray}
Here, it should be emphasized that the definition of the mean arrival time
used in Eq.(2) is {\it not} a unique result within standard quantum mechanics.
We also note that ${\bf J}({\bf x},t)$ can be negative, hence one needs to
take the modulus sign in order to use the above definition. However, the
Bohmian model of quantum mechanics in terms of the causal trajectories of
individual particles leads to the above expression for the mean arrival
time\cite{leavens2} in a rather conceptually elegant way.

The quantum probability current interpreted as the streamlines of a
conserved flux has been used for studying the tunneling times
of Dirac electrons\cite{challinor}. However, it is easily seen that 
in non-relativistic quantum mechanics the form of the probability
current density is ${\it not~unique}$, a point which has already been
discussed by a number of authors\cite{deotto,peter,finkelstein}. 
If we replace ${\bf J}$ by ${\bf J}^{\prime}$ in Eq.(1) where 
${\bf J}^{\prime}$=${\bf J}$+$\delta$${\bf J}$, 
with ${\bf \nabla}$.$\delta$${\bf J}$=0, ${\bf J}^{\prime}$ satisfies the 
same probability conservation given by Eq.(1).
Then this new current density ${\bf J}^{\prime}$ can lead to a different 
distribution function for the arrival
time\cite{finkelstein}. Hence the question arises ${\it how}$ one can uniquely
fix the arrival time distribution via the quantum probability current in the 
regime of non-relativistic quantum mechanics? 

In order to address the above question, we take a vital clue from
the interesting result Holland\cite{holland} had shown in the context of
analysing the uniqueness of the Bohmian model of quantum mechanics,
viz. that the Dirac equation implies a ${\it unique}$ expression for
the probability current density for spin 1/2 particles in the non-relativistic
regime. In Section II we highlight the feature that the uniqueness
of the probability current density is a ${\it generic}$ consequence
of ${\it any}$ relativistic equation of quantum dynamics. In Section
III, the particular case of spin dependent probability current density
as derived from the Dirac equation is discussed in detail. Subsequently,
in Section IV, using the non-relativistic limit of the Dirac current
density, we compute the effect of spin on the arrival time distribution
of free particles for an initial Gaussian wave packet. Such a line
of investigation has not been explored sufficiently; to the best of our
knowledge, it seems that only Leavens\cite{leavens3} has studied this specifically 
in terms of the Bohmian causal model of spin-1/2 particles.

\section{Uniqueness of the probability current density for any relativistic wave equation} 
 
 The current obtained from any consistent relativistic quantum wave equation will
have to satisfy the ${\it covariant}$ form of the continuity equation $\partial_\mu  j^\mu=0$,
where the zeroth component of $j^\mu$ will be associated with the probability density. Now,
let us replace $j^\mu$ by $\overline { j}^\mu$ which should again be conserved, i.e. 
$\partial_\mu \overline{ j}^\mu=0$, where $\overline{ j}^\mu= j^\mu + a^\mu$,~ $a^\mu$
being an arbitrary 4-vector. But then the zeroth component $\overline{j}^0$ will have to
reproduce the same probability density $j^0$, and hence $a^0$=0. 
This current as seen from another Lorentz frame is ${j^\mu}^{\prime}$=$j^{\mu}$+${a^\mu}^{\prime}$.
Then in this frame ${j^0}^{\prime}$=$j^0$+${a^0}^{\prime}$, and again from the previous argument 
${a^0}^{\prime}$=0. But we know that the only 4-vector whose fourth
component vanishes in all frames is the null vector. Hence $a^\mu$=0.
Thus, for any consistent relativistic quantum wave equation, satisfying the covariant 
form of the continuity equation, the relativistic current is uniquely fixed. 
Unique expressions for the conserved currents have been explicitly derived
by Holland\cite{holland2} for the Dirac equation, the Klein-Gordon equation,
and also for the coupled Maxwell-Dirac equations.

Now, an interesting point is that this uniqueness 
will be preserved in the non-relativistic regime. Hence, given ${\it any}$ relativistic 
wave equation, one can calculate the unique form of the current which can be applied 
in the non-relativistic regime. Thus using the (normalized) modulus of the probability 
current density as the arrival time distribution, if one calculates the mean 
arrival time, it can be used to empirically test any 
consistent relativistic wave equation such as the relativistic Kemmer equation\cite{kemmer} for the massive spin 0 and spin 1 bosons. Of late, the unique form of the probability 
current density expressions has been derived in the non-relativistic limit of the 
relativistic Kemmer equation for spin-0 and spin-1 particles\cite{baere}.
This general scheme for testing relativistic quantum wave equation in terms 
of the arrival time distribution is not contingent 
on any specific form of the relativistic wave equation. However, in the following 
detailed study we specifically use the Dirac equation for spin-1/2 particles. 

\section{spin dependent effect on the arrival time distribution using dirac equation}

The Dirac equation for a ${\it free}$ ${\it particle}$ is 
\begin{eqnarray}
i\hbar\frac{\partial \psi}{\partial t} = \Biggl(\frac{\hbar c}{i}\hskip 0.2cm {\alpha}_i \frac{\partial}{\partial x_i}+\beta m_0 c^2 \Biggr)\psi
\end{eqnarray}
\[\alpha_i = \left (\begin{array}{cc}
0 & \sigma_i\\
\sigma_i & 0 \\
\end{array}\right),\beta = \left (\begin{array}{cc}
I & 0\\
0 & -I \\
\end{array}\right),\psi = \left (\begin{array}{c}
\psi_1\\
\psi_2\\
\end{array}\right)\]

$\psi$ is a four component column matrix and $\sigma_i$ are the Pauli matrices.  
Choosing a representation where $\psi_1$ and $\psi_2$ are two component spinors,
one gets two coupled equations

\begin{equation}
\frac{\partial \psi_1}{\partial t}= -c \sigma_i \frac{\partial \psi_2}{\partial x^i}-\frac{im_0
c^2}{\hbar} \psi_1
\end{equation}
\begin{equation}
\frac{\partial \psi_2}{\partial t}= -c \sigma_i \frac{\partial \psi_1}{\partial x^i}+\frac{im_0
c^2}{\hbar} \psi_2
\end{equation}

Combining Eqs.(4) and (5) one gets
\begin{equation}
\frac{\partial}{\partial t}({\psi_1}^\dagger \psi_1)=-c {\psi_1}^\dagger \sigma_i
\frac{\partial \psi_2}{\partial x^i}-c \frac{\partial {\psi_2}^\dagger}{\partial x^i}\sigma_i \psi_1
\end{equation}
For positive energies, one  can take
$\psi_2 \propto exp(-iEt/\hbar)$. In the non-relativistic limit, E is the rest
mass energy and $E+m_0 c^2=(m+m_0)c^2\cong2m_0 c^2$. Thus using Eq.(5) one
can write
\begin{equation}
\psi_2=-\frac{i\hbar c}{(E+m_0 c^2)}\sigma_i \frac{\partial \psi_1}{\partial x^i}=-\frac{i\hbar}{2m_0 c}\sigma_i \frac{\partial \psi_1}{\partial x^i}
\end{equation}
Putting this value of $\psi_2$ in Eq.(6), one  gets
\begin{equation}
\frac{\partial \rho}{\partial t}+{\bf  \nabla} . {\bf J} =0
\end{equation}
where $ {\bf J}$ is the Dirac current in the non-relativistic limit that can 
be decomposed into two terms as was shown by Holland\cite{holland,holland2}, as

\begin{eqnarray} 
{\bf J}=-\frac{i\hbar}{2m}\left[{\psi_1}^\dagger {\bf \sigma}({\bf \sigma}. {\bf \nabla})
\psi_1-({\bf \nabla}{\psi_1}^\dagger.{\bf \sigma}){\bf \sigma} \psi_1  \right]
\\
=-\frac{i\hbar}{2m}\left[{\psi_1}^\dagger ({\bf \nabla} \psi_1)
-({\bf \nabla} {\psi_1}^\dagger)\psi_1 \right]
+ \frac{\hbar}{2m} {\bf \nabla} \times ({\psi_1}^\dagger
{\bf \sigma} \psi_1)\nonumber 
\end{eqnarray}
and $\rho={\psi_1}^\dagger \psi_1$. \hskip 0.2cm  $\psi_1$ is a two component spinor 
which can be written for a particle in a spin eigenstate as
\begin{eqnarray}
\psi_1&=&\psi(x,t)\chi 
\equiv \left[R(x,t)\hskip 0.1cm {\rm exp}\left(\frac{iS(x,t)}{\hbar}\right)\right]\chi
\end{eqnarray}
  
Here $\psi(x,t)$ is the Schroedinger wavefunction and $\chi$
is a spin eigenstate. Putting this form of $\psi_1$ in the  expression for current in Eq.(9) one gets
\begin{eqnarray}
{\bf J}&=&\frac{1}{m} \rho {\bf \nabla} S +\frac{1}{m}\left({\bf \nabla} \rho \times {\bf s}\right)\nonumber \\
\equiv {{\bf J}}_{i} + { {\bf J}}_{s}
\end{eqnarray}
with
\begin{eqnarray*}
{\bf s}=(\hbar / 2)\chi^\dagger {\bf \sigma} \chi,\hskip 1.0cm  \rho=R^2,\hskip 1.0cm \chi^\dagger
\chi=1
\end{eqnarray*}

The first term (${{\bf J}}_{i}$) in Eq.(11) is independent of spin,
while the second term (${{\bf J}}_{s}$) is the contribution of the spin 
of a free particle to the unique conserved vector current in the non-relativistic
limit. It is then clear that the mean arrival time given
by Eq.(2) can be computed by using the unique expression for ${\bf J}$ in Eq.(11). Thus
we can obtain a spin-dependent contribution in the expression for the mean time of 
arrival for free particles,
which could be an experimentally 
measurable quantity. On the other hand, by ignoring the spin-dependent term one would obtain the 
mean arrival time given by
\begin{eqnarray}
\bar {\bf \tau}_i=\frac{\int^{\infty}_{0} |{\bf J}_{i}| t  dt}
{\int^{\infty}_{0} |{\bf J}_{i}|  dt}
\end{eqnarray} 
In the following section IV we study the situations where the difference
between the actual magnitudes of $\bar {\bf \tau}$ and $\bar {\bf \tau}_i$ is 
significant, thereby enhancing the feasibility of experimentally testing the
specific spin-dependent effect.

\section{The computed effects on the arrival time distribution}
We consider a freely evolving Gaussian wave packet in the two separate cases (A and B)
corresponding to an initially symmetric and an asymmetric wave packet respectively.

{\bf Case A}: Symmetric wave packet

Let us consider a Gaussian wave packet for a  free spin $1/2$ particle 
of mass $m$ centered at the point $x=0$, $y=0$,and  $z=0$. We choose the 
spin to be
directed along the $z$-axis, i.e., (${\bf s}=\frac{1}{2}{\hat z}$).  
\begin{eqnarray}
\psi({\bf x},t=0)=\frac{1}{(2\pi {\sigma_0}^2)^{3/4}}
{\rm exp}(i{\bf k}.{\bf x})
{\rm exp}\left(-\frac{{\bf x}^2}{4 {\sigma_0}^2}\right) 
\end{eqnarray}                                                         
The time evolved wave function can be
written as
\begin{eqnarray} 
\psi({\bf x},t)=R({\bf x},t) {\rm exp}\left[\frac{i S({\bf x},t)}{\hbar} \right]
\end{eqnarray}
where 
\begin{eqnarray}
R({\bf x},t) = \bigl(2\pi\sigma^2\bigr)^{-3/4}{\rm exp}\Biggl[-\frac{({\bf x}
- {\bf u}t)^2}{4\sigma^2}\Biggr]
\end{eqnarray} 
and
\begin{eqnarray} 
S({\bf x},t)&=&-\frac{3\hbar}{2}\tan^{-1}\left(\frac{\hbar t}{2m {\sigma_0}^2}\right)\\
&&+ m{\bf u}.({\bf x} - {1\over 2}{\bf u}t) + {({\bf x} - {\bf u}t)^2 \hbar^2
t \over 8m\sigma_0^2 \sigma^2}\nonumber
\end{eqnarray} 
with (${\bf u}= \hbar {\bf k}/m$) the initial group velocity taken along
the $x$-axis, and
\begin{eqnarray}
\sigma = \sigma_0\Biggl[1 + {\hbar^2 t^2 \over 4m^2\sigma_0^4}\Biggr]^{1/2}
\end{eqnarray}
The total current density can be calculated using Eq.(11) to be 
(we set $m=1$, $\hbar$=1)
\begin{eqnarray}
{\bf J} &=& \rho\Biggl[\Biggl(u + {(x-ut)t\over 4\sigma_0^2\sigma^2}\Biggr)
{\bf\hat x} + \Biggl({yt\over 4\sigma_0^2\sigma^2}\Biggr){\bf\hat y} +
\Biggl({zt\over 4\sigma_0^2\sigma^2}\Biggr){\bf\hat z}\Biggr]\nonumber \\
&+& \rho\Biggl[-\Biggl({y\over 2\sigma^2}\Biggr){\bf\hat x} + {(x-ut)\over
2\sigma^2}{\bf\hat y}\Biggr] 
\end{eqnarray}
where the contribution of spin is contained in the second term only.

We can now compute $\bar {\bf \tau}$ and $\bar {\bf \tau}_i$ numerically by 
substituting Eq.(18) 
in Eqs.(2) and (12) respectively. It is instructive
to examine the behaviour of the contribution of spin-dependent term towards
the mean arrival time. For this purpose, we define a quantity
\begin{equation}
\bar {\bf \tau}_s=\frac{\int^{\infty}_{0} |{\bf J}_{s}| t  dt}
{\int^{\infty}_{0} |{\bf J}_{s}|  dt}
\end{equation} 
We first compute $\bar {\bf \tau}_s$ for a range of the initial velocity $u$
in units of m=1, and $\hbar$=1. We find that the spin of a free particle
contributes towards altering its mean arival time for a wide range of
initial velocities. This feature holds generally, except for very small
magnitudes of velocity where the spin-dependent contribution may be
negligible depending on the location of the detector vis-a-vis the
direction of the initial group velocity ${\bf u}$. This feature is shown in
Figure.1 where  we plot the variation of $\bar {\bf \tau}_s$ with
$u$. The initial wave packet is peaked at the origin with $\sigma_0=0.01$. 
The detector position is chosen at ($x=1, y=1, z=1$). We also find
that the difference of magnitude between $\bar {\bf \tau}$ and $\bar {\bf \tau}_i$ 
can be
increased by choosing asymmetric detector positions as well as asymmetric
spread for the initial wave packet, an example of which we highlight below.
 
\begin{figure}
\begin{center}
\centerline{\epsfig{file=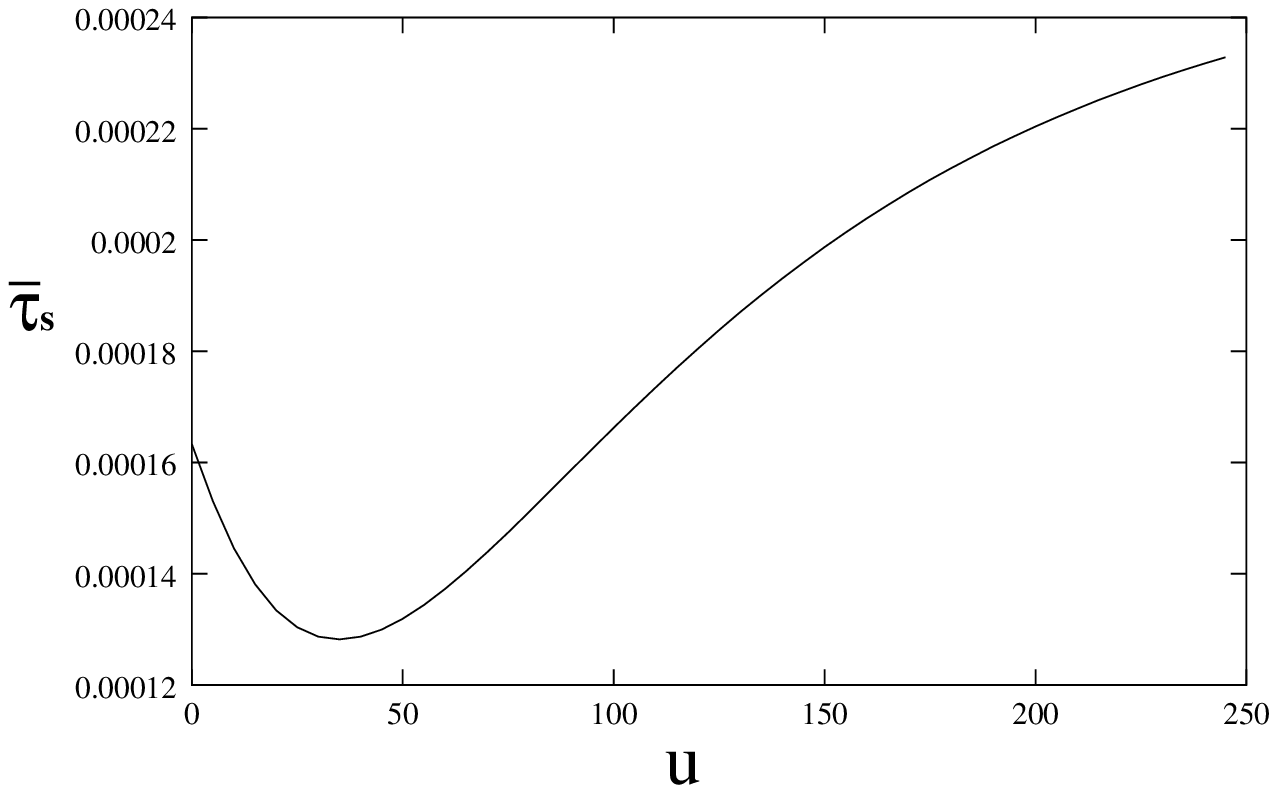,height=2.7in,angle=0}}
\narrowtext{\caption{The spin-dependent contribution to the mean arrival
time computed at the point x=1, y=1, z=1 is plotted against the
initial group velocity of the packet along the x-axis.}} 
\end{center}
\end{figure}                                                                      

\noindent
{\bf Case B}: Asymmetric wave packet

We consider an initial free particle wave packet in
three dimensions which is centered at the point
$x=-x_1$, $y=0$, $z=0$.
\begin{eqnarray}
&&\psi(x,y,z,t=0)=\left(\frac{1}{\pi^3 a^2 b^2 c^2}\right)^{1/4}{\rm exp}(ikx) 
\nonumber \\
&&{\rm exp}\left[-\frac{(x+x_1)^2}{2 a^2} \right] 
{\rm exp}\left[-\frac{(y)^2}{2 b^2} \right] 
{\rm exp}\left[-\frac{(z)^2}{2 c^2} \right]
\end{eqnarray}
where $a, b, c$  are positive constants. (Such a form for the wave function 
was considered by Finkelstein\cite{finkelstein} in the context of arrival time distributions.)
The particle is given an initial
velocity in the $x$ direction represented by $u=\frac{\hbar k}{m}$.
The time evolved wave function is given by
\begin{eqnarray}
&&\psi(x,y,z,t)=\left(\frac{a^2 b^2 c^2}{\pi^3 }\right)^{1/4}
\frac{{\rm exp}[i(kx-{k^2 t}/2)]}{\alpha \beta \gamma} \nonumber \\
&&{\rm exp}\left[-\frac{(x+x_1-k t)^2}{2 \alpha^2} \right] {\rm exp}\left[-\frac{y^2}{2 \beta^2} \right] {\rm exp}\left[-\frac{z^2}{2 \gamma^2} \right]
\end{eqnarray}\\
where
 $\alpha=(a^2 + it)^{1/2}$; $\beta=(b^2 +  it)^{1/2}$; $\gamma=(c^2 +it)^{1/2}$.

Writing the wave function as
\begin{eqnarray}
\psi(x,y,z,t)=R(x,y,z,t) {\rm exp}\left[\frac{i S(x,y,z,t)}{\hbar} \right]
\end{eqnarray}
one obtains
\begin{eqnarray}
&&R(x,y,z,t)=\left(\frac{a^2 b^2 c^2}{\pi^3 }\right)^{1/4}
\frac{1}{(p^2 + q^2)^{1/4}}\nonumber\\
&&{\rm exp}\left[-\frac{a^2(x+x_1-k t)^2}{2(a^4 +t^2)} \right] {\rm exp}\left[-\frac{b^2 y^2}{2(b^4 +
t^2)} \right] \nonumber\\
&&{\rm exp}\left[-\frac{c^2 z^2}{2(c^4 +t^2)} \right]
\end{eqnarray} 
and
\begin{eqnarray}
&&S(x,y,z,t)=\hbar k x - \frac{\hbar k^2 t}{2}
- \frac{\hbar}{2}\tan^{-1}(q/p)\nonumber\\
&& + \frac{\hbar t (x+x_1-kt)^2}{2(a^4+t^2)} +\frac{\hbar t y^2}{2(b^4+t^2)}+ \frac{\hbar
t z^2}{2(c^4+t^2)}
\end{eqnarray} 
with
\begin{eqnarray}
p=a^2 b^2 c^2- a^2 t^2 -b^2 t^2 -c^2 t^2\nonumber \\
q=a^2 b^2 t + a^2 c^2 t + b^2 c^2 t - t^3
\end{eqnarray}

Considering again a spin-1/2 particle with spin directed along z-axis
(${\bf s}=\frac{1}{2}{\hat z}$), the total
current density defined in Eq.(11) is given by (in units of $\hbar=1=m$)
\begin{eqnarray}
{\bf J}&=&\rho \left[\left(u+\frac{(x+x_1-ut)t}{(a^4+t^2)}\right){\bf\hat x}+\frac{y t}{(b^4+t^2)}{\bf\hat y}+\frac{z t}{(c^4+t^2)}{\bf\hat z}\right] 
\nonumber\\
&&+\rho\left[-\frac{b^2 y}{(b^4+t^2)}{\bf\hat x} + \frac{a^2(x+x_1-ut)}{(a^4+t^2)}{\bf\hat y}\right]
\end{eqnarray}
where the second term represents the spin-dependent contribution to the current.

\begin{figure}
\begin{center}
\centerline{\epsfig{file=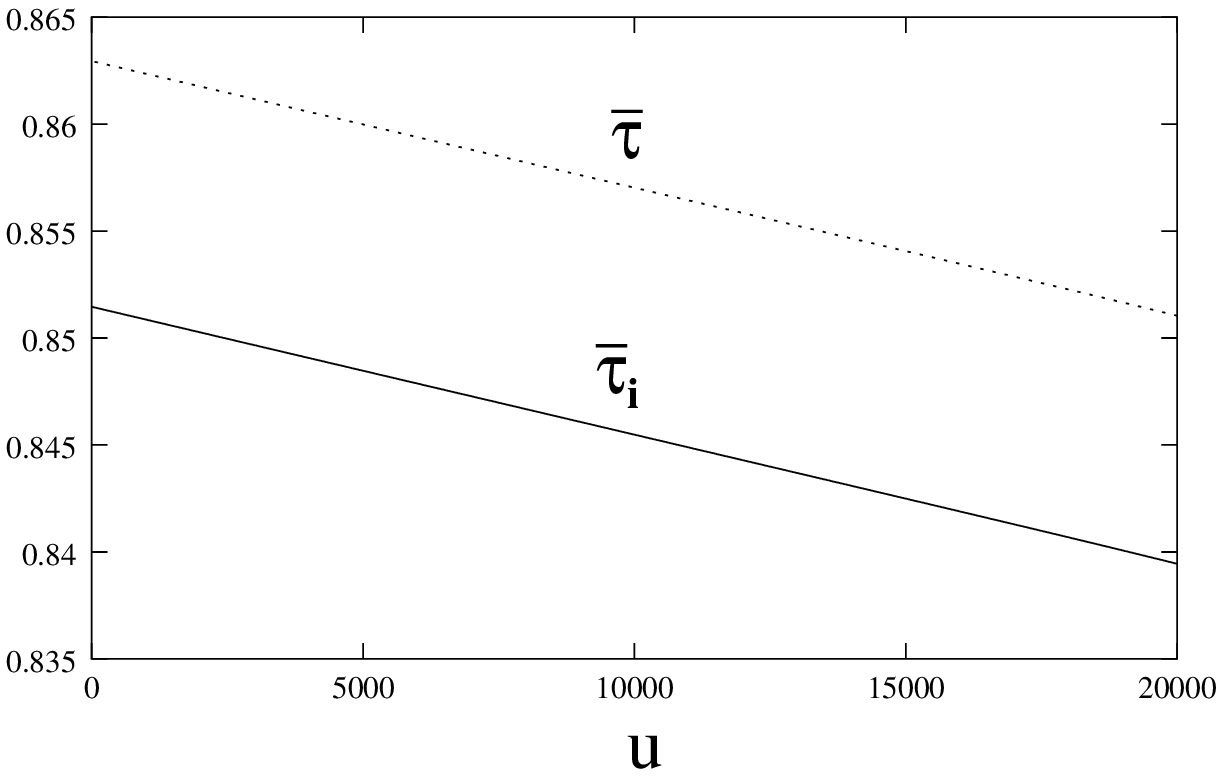,height=2.7in,angle=0}}
\narrowtext{\caption{The mean arrival times $\bar {\bf \tau}$ (upper curve) 
and $\bar {\bf \tau}_i$ (lower curve)
computed at the point $x=1, y=2, z=1$ are plotted against the
initial group velocity of the packet along the $x$-axis. and 
$a=0.001, b=0.4, c=0.01$, $x_1=0$.}}
\end{center}
\end{figure}  

We compute numerically the arrival times $\bar {\bf \tau}$ and $\bar {\bf \tau}_i$. Figure.2 
shows the variation of $\bar {\bf \tau}$ and $\bar {\bf \tau}_i$ with
the initial group velocities (u) of the wave packet. Here we choose the parameters 
as $x_1=0, a=0.001, b=0.4, c=0.01$. Accordingly  the mean arrival time is computed
at the position $x=1.0, y=2.0, z=1.0$. One sees that the difference in the 
magnitudes of $\bar {\bf \tau}$ 
and $\bar {\bf \tau}_i$ can suitably be enhanced by a judicious choice 
of asymmetric initial spreads and detector positions.
\vskip 0.2 cm

\section{concluding remarks}

Let us now summarise the salient features of our scheme. For measuring
the spin of a particle, it is usually subjected to an external field, like
in a Stern-Gerlach apparatus. But the scheme we have suggested would enable
to detect, a spin-dependent effect ${\it without}$ using any external 
field. Such an observable effect thus highlights the feature that
the spin of a particle is an ${\it intrinsic}$ property and is 
\emph{not} contingent on the presence of an external field. As demonstrated
in this paper, the spin-dependent term in the Dirac probability current
density contributes significantly to the computed mean arrival time
for a range of suitably chosen parameters of the Gaussian wave packet. Thus
if the arrival time distribution can be measured, this predicted
spin-dependent effect would be empirically verifiable.

Another way of perceiving the significance of such an effect is as follows.
The dynamical properties of free particles like position, momentum, and energy
can of course be measured. However, one cannot usually measure the \emph{static} 
or \emph{innate} particle properties such as charge without using any external 
field. Nevertheless, the scheme we have proposed shows that the magnitude of total
spin  can be measured without subjecting the particle to an external field.

Another interesting implication of the measurement of the spin dependent arrival 
times for free particles could be to view this as implying a fundamental 
difference between the magnitude of total spin of a particle and its other 
static properties
such as mass and charge in the following sense. The property of spin seems
to be crucially contingent on the fundamentally relativistic nature of the
dynamical equation of the wave function of the particle so that the wave
function is essentially $4$-component, or $2$-component in the non-relativistic
limit.

Here we would like to stress that the spin-dependent term which contributes
significantly to the arrival time distribution we have computed in the nonrelativistic
regime originates from the relativistic Dirac equation and hence this
provides a rather rare example of an empirically detectable manifestation
of a relativistic dynamical equation in the nonrelativistic regime, an effect
which \emph{cannot} be derived uniquely from the Schroedinger dynamics.

A further line of investigation as an offshoot of this paper could
be to explore the possibilities of using the relativistic quantum
mechanical wave equations of particles with spins other than spin 1/2 (such
as using the Kemmer equation\cite{kemmer,baere} for spin 0 and spin 1 particles)
in order to compute the spin-dependent terms in the probability current
densities and their effects on the arrival time distribution. Such
a study seems worthwhile because then the arrival time distribution
may provide a means of checking the validity of the various suggested 
relativistic quantum mechanical equations which have
otherwise eluded any empirical verification.

\vskip 0.2in

We thank Peter Holland for some stimulating suggestions and for his discussions with
ASM.  We are also grateful to thank Rick Leavens for his very helpful comments on 
the initial version of this paper. This research work of DH is supported by the 
Jawaharlal Nehru Fellowship. MA acknowledges the Junior Research Fellowship from 
CSIR, India.

\end{multicols}

\end{document}